\documentclass[aip,jcp,twocolumn,superscriptaddress,showpacs,showkeys,reprint,floatfix]{revtex4-1}

\usepackage{amssymb}
\usepackage{amsmath}
\usepackage{graphicx}
\usepackage{hyperref}   
\usepackage{floatrow}
\usepackage{color}

\newcommand*{\citen}[1]{%
  \begingroup
    \romannumeral-`\x 
    \setcitestyle{numbers}%
    \cite{#1}%
  \endgroup   
}

\begin{document}

\title{Connectedness percolation of hard deformed rods}

\author{Tara Drwenski}
\email{t.m.drwenski@uu.nl}
\affiliation{Institute for Theoretical Physics, Center for Extreme Matter and Emergent Phenomena, Utrecht University, Princetonplein 5, 3584 CC Utrecht, The Netherlands}

\author{Simone Dussi} 
\affiliation{Soft Condensed Matter, Debye Institute for Nanomaterials Science, Utrecht University, Princetonplein 5,
3584 CC Utrecht, The Netherlands}

\author{Marjolein Dijkstra} 
\affiliation{Soft Condensed Matter, Debye Institute for Nanomaterials Science, Utrecht University, Princetonplein 5,
3584 CC Utrecht, The Netherlands}

\author{Ren\'{e} van Roij} 
\affiliation{Institute for Theoretical Physics, Center for Extreme Matter and Emergent Phenomena, Utrecht University, Princetonplein 5, 3584 CC Utrecht, The Netherlands}

\author{Paul van der Schoot} 
\email{p.p.a.m.v.d.schoot@tue.nl}
\affiliation{Institute for Theoretical Physics, Center for Extreme Matter and Emergent Phenomena, Utrecht University, Princetonplein 5, 3584 CC Utrecht, The Netherlands}
\affiliation{Theory of Polymers and Soft Matter, Eindhoven University of Technology, P.O. Box 513, 5600 MB Eindhoven, The Netherlands}

\date{\today}

\begin{abstract}
Nanofiller particles, such as carbon nanotubes or metal wires, are used in functional polymer composites to make them conduct electricity. They are often not perfectly straight cylinders, but may be tortuous or exhibit kinks. Therefore we investigate the effect of shape deformations of the rodlike nanofillers on the geometric percolation threshold of the dispersion. We do this by using connectedness percolation theory within a Parsons-Lee type of approximation, in combination with Monte Carlo integration for the average overlap volume in the isotropic fluid phase. We find that a deviation from a perfect rodlike shape has very little effect on the percolation threshold, unless the particles are strongly deformed. This demonstrates that idealized rod models are useful even for nanofillers that superficially seem imperfect. In addition, we show that for small or moderate rod deformations, the universal scaling of the percolation threshold is only weakly affected by the precise particle shape.
\end{abstract}

\maketitle

\section{Introduction}\label{sect:intro}

Nanocomposites of carbon nanotubes or metallic wires dispersed in plastics are seen to be promising replacements of indium tin-oxide for transparent electrodes.\cite{hecht2011,mutiso2015} Opto-electronic applications of this material require as low as possible percolation thresholds, to keep the materials transparent. The percolation threshold, the critical filler loading required to get significant electrical conduction, depends crucially on the formulation and processing of the composite. It is not surprising then that a significant amount of effort has been invested and continues to be invested in understanding what factors precisely control the percolation threshold.\cite{mutiso2015}

Continuum percolation theory and computer simulations of highly idealized models of the elongated filler particles, usually modeled as hard rods or ellipsoids, indicate that the filler fraction at the percolation threshold should be of the order of the inverse aspect ratio of the particles.\cite{mutiso2015} Similar models have been invoked to study the impact of length and width polydispersity,\cite{kyrylyuk2008,otten2009,otten2011,mutiso2012,nigro2013,meyer2015} attractive interactions,\cite{kyrylyuk2008} alignment,\cite{white2009,otten2012} etc. While very informative, the question arises how accurate these idealized models are. Indeed, carbon nanotubes are often not straight cylinders but instead quite tortuous or riddled with kink defects.\cite{dalmas2006,lucas2009} The same is true for other types of conductive filler particles. However, little theoretical effort has been put into studying the applicability of these perfect rod models to systems with shape defects.

In this paper, we investigate the impact of the precise shape of the rods upon the percolation threshold. For this purpose, we apply connectedness percolation theory to kinked and bent rods. Here we vary the aspect ratio, the kink location and angle, and the curvature. We find that the main contributing factor in determining the percolation threshold is the aspect ratio, not the precise shape of the particle, unless it is extremely deformed. This implies that idealized models are indeed useful in an experimental context. We also find that the universal scaling of the percolation threshold with particle length and connectivity range is only very weakly affected by the particle shape.

The remainder of this paper is structured as follows. In Sec.~\ref{sect:method}, we present the methodology that we base our calculations on. We use connectedness percolation theory within the second-virial approximation, augmented by the Parsons-Lee correction in order to account for finite-size effects. We use Monte Carlo integration to calculate the overlap volumes of the particles. In Sec.~\ref{sect:results}, we present our results and we summarize our findings in Sec.~\ref{sect:conclusions}.

\section{Method}\label{sect:method}

Here we study the size of clusters of connected particles, where we define two particles as connected if their surface-to-surface distance is less than a certain connectedness criterion (connectedness range) $\Delta$. This connectedness criterion is related to the electron tunneling distance and depends on the nanofiller properties as well as the dielectric properties of the medium.\cite{ambrosetti2010,kyrylyuk2008} Using connectedness percolation theory,\cite{hill1955,coniglio1977}  we study the average cluster size of connected particles. Specifically, we are interested in the percolation threshold, that is, the lowest density at which the average cluster size diverges.

For completeness and clarity, we now give the full derivation of the percolation threshold. Letting $n_k$ denote the number of clusters of $k=1,2,\ldots$ particles, then the probability of a particle being in a cluster of size $k$ is simply $s_k = k n_k/N$, where $N = \sum_k k n_k$ is the total number of particles.\cite{torquato2002} Then the weight-averaged number of particles in a cluster is defined as $S = \sum_k k s_k = \sum_k k^2 n_k/N$. This can be rewritten as $S= \sum_k (k n_k + k(k-1)n_k)/N = 1+ 2N_c/N$, where in the last step we defined $N_c = \sum_k k(k-1)n_k/2$, which is the number of pairs of particles within the same cluster.\cite{torquato2002} The density at which $S$ diverges is the percolation threshold, and in addition $S$ can be probed indirectly by measuring the frequency-dependent dielectric response, which has a sharp peak at the percolation threshold.\cite{vigolo2005} Below we calculate $N_c$, and thus $S$.

Now we consider clusters composed of rigid, non-spherical particles. The orientation of such particles can be given by three Euler angles $\Omega = (\alpha,\beta,\gamma)$. Assuming a uniform spatial distribution of particles with number density $\rho$, the orientation distribution function $\psi(\Omega)$ is defined so that the probability to find a particle with an orientation in the interval $d\Omega$ is given by $\psi(\Omega) d\Omega$, with the normalization constraint that $\int d \Omega \, \psi(\Omega) = \int_0^{2\pi} d\alpha \,\int _0^{\pi}  d\beta \sin \beta \, \int_0^{2\pi} d\gamma  \, \psi(\Omega) = 1$. The orientational average is denoted $\langle \ldots \rangle_\Omega = \int d \Omega \, \ldots \psi(\Omega)$.

The pair connectedness function $P$ is defined such that $\rho^2 P(\mathbf{r}_1, \mathbf{r}_2,\Omega_1,\Omega_2) \psi(\Omega_1) \psi(\Omega_2) d\mathbf{r}_1 d\mathbf{r}_2 d\Omega_1 d\Omega_2 $ is the probability of finding a particle in volume $d\mathbf{r}_1$ with orientation in $d\Omega_1$ and a second particle in volume $d\mathbf{r}_2$ with orientation in $d\Omega_2$, given that the two particles are in the same cluster.\cite{coniglio1977} From this definition, it follows that 
\begin{eqnarray}
	N_c &=& \frac{\rho^2}{2} \int  d\mathbf{r}_1 \int d\mathbf{r}_2  \langle \langle P(\mathbf{r}_1, \mathbf{r}_2,\Omega_1,\Omega_2) \rangle_{\Omega_1} \rangle_{\Omega_2}\nonumber\\
		&=&	\frac{\rho N}{2} \int  d\mathbf{r}_{12}  \langle \langle P(\mathbf{r}_{12} ,\Omega_1,\Omega_2) \rangle_{\Omega_1} \rangle_{\Omega_2},\label{eq:nc2}
\end{eqnarray}
where the one-half prefactor avoids double counting and in the second line of Eq.~\eqref{eq:nc2} we assume translational invariance with $\mathbf{r}_{12}=\mathbf{r}_1-\mathbf{r}_2$.  It follows that the weight-averaged cluster size can be written as
\begin{equation}
	S = \lim_{\mathbf{q} \to 0 } S(\mathbf{q}),
\end{equation}
with
\begin{equation}
	S(\mathbf{q}) =1+ \rho  \langle \langle \hat{P}(\mathbf{q}, \Omega_1, \Omega_2) \rangle_{\Omega_1} \rangle_{\Omega_2} ,
\end{equation}
where we denote the Fourier transform of an arbitrary function $f$ by $\hat{f}(\mathbf{q}) = \int d \mathbf{r} f(\mathbf{r}) \exp(i \mathbf{q} \cdot \mathbf{r})$.

The Fourier transform of the pair connectedness function obeys the connectedness Ornstein-Zernike equation, given by\cite{coniglio1977}
\begin{eqnarray}\label{eq:OZ}
	\hat{P}(\mathbf{q},\Omega_1,\Omega_2) &=& \hat{C}^+(\mathbf{q},\Omega_1,\Omega_2) \nonumber \\
	&\quad&+ \rho \langle \hat{C}^+(\mathbf{q},\Omega_1,\Omega_3)\hat{P}(\mathbf{q},\Omega_3,\Omega_2) \rangle_{\Omega_3},
\end{eqnarray}
with $\hat{C}^+(\mathbf{q},\Omega_1,\Omega_2)$ the spatial Fourier transform of the direct pair connectedness function which measures short-range correlations. Given a closure for $\hat{C}^+$, we can calculate $\hat{P}(\mathbf{q},\Omega_1,\Omega_2)$ and thus $S(\mathbf{q})$. 

In this paper we only consider percolation in the isotropic phase, where all orientations are equally probable and so $\psi(\Omega)=1/(8\pi^2)$. Due to the global rotational invariance of the system and symmetry under particle exchange, the pair connectedness function $\hat{P}$ has the following properties
\begin{eqnarray}\label{eq:properties}
	&\hat{P}&(\mathbf{q},\Omega_1,\Omega_2) = \hat{P}(\mathbf{q},\Omega_{12}) = \hat{P}(q,\Omega_{12})= \hat{P}(q,\Omega_{21}),
\end{eqnarray}
where $q = |\mathbf{q}|$ and $\Omega_{12}=\Omega_1^{-1}\Omega_2$ denotes the relative orientation between particle $1$ and particle $2$. Analogous properties hold for $\hat{C}^+$.

Using the properties in Eq.~\eqref{eq:properties} and integrating both sides in Eq.~\eqref{eq:OZ} over $\Omega_{2}$ gives
\begin{eqnarray}\label{eq:OZ2}
	\langle \hat{P}(q,\Omega_{12}) \rangle_{\Omega_{2}} &=& \langle \hat{C}^+(q,\Omega_{12}) \rangle_{\Omega_{2}} \nonumber \\
	&\quad&+ \rho \langle \hat{C}^+(q,\Omega_{13}) \, \langle \hat{P}(q,\Omega_{32})\rangle_{\Omega_{2}} \,\rangle_{\Omega_{3}}.
\end{eqnarray}
Consider the second term on the right-hand side of Eq.~\eqref{eq:OZ2}. By a measure-invariant change of variables $\Omega_2 \to \Omega_3^{-1}\Omega_2 = \Omega_{32}$, we find that $\langle \hat{P}(q,\Omega_{32}) \rangle_{\Omega_{2}} = \langle \hat{P}(q,\Omega_{32}) \rangle_{\Omega_{32}}$. By subsequently performing similar changes of variables on the remaining integrals in Eq.~\eqref{eq:OZ2}, we find
\begin{eqnarray}\label{eq:OZ3}
	\langle \hat{P}(q,\Omega_{12}) \rangle_{\Omega_{12}} &=& \langle \hat{C}^+(q,\Omega_{12}) \rangle_{\Omega_{12}}  \nonumber \\
	&\quad&+ \rho \langle \hat{C}^+(q,\Omega_{13}) \rangle_{\Omega_{13}} \langle \hat{P}(q,\Omega_{32}) \rangle_{\Omega_{32}} 
\end{eqnarray}
which can be solved as 
	\begin{equation}\label{eq:OZ4}
		\langle \hat{P}(q,\Omega) \rangle_\Omega = \frac{\langle \hat{C}^+(q,\Omega) \rangle_\Omega}{1- \rho \langle \hat{C}^+(q,\Omega) \rangle_\Omega }.
	\end{equation}
Therefore the weight-averaged cluster size obeys
	\begin{equation}\label{eq:S}
			S = \frac{1}{1- \rho \lim_{q \to 0}\langle \hat{C}^+(q,\Omega) \rangle_\Omega }.
	\end{equation}
The percolation threshold is defined as the density at which Eq.~\eqref{eq:S} diverges, i.e.,
	\begin{equation}\label{eq:percDensity}
		\rho_P = \dfrac{1}{  \lim_{q \to 0} \langle \hat{C}^+(q,\Omega) \rangle_\Omega }.
	\end{equation}

For hard spherocylinders with length $L$ much larger than diameter $D$, the second-virial approximation is very accurate, and in fact becomes exact as $L/D \to \infty$.\cite{onsager1949,otten2011} The closure is then given by $\hat{C}^+(q,\Omega_{12}) = \hat{f}^+(q,\Omega_{12})$, where the Fourier transform of $\hat{f}$ is the connectedness Mayer function $f^+(\mathbf{r},\Omega_{12}) = \exp(-\beta U^+(\mathbf{r},\Omega_{12}))$, with $\beta$ the inverse thermal energy, and $U^+$ the connectedness pair potential,\cite{coniglio1977,bug1986} which can be written as
\begin{equation}
\label{eq:potential}
 \beta  U^+(\mathbf{r},\Omega_{12})= \left\{
     \begin{array}{cl}
       0, & \text{1 and 2 are connected;}  \\
       \infty, & \text{otherwise} ,
     \end{array}
   \right. 
\end{equation}
where we adopt the so-called core-shell model.\cite{berhan2007a} This consists of defining two particles as connected if their shortest surface-to-surface distance is less than connectedness criterion $\Delta$, i.e., their shells of diameter $D+\Delta$ overlap, but an overlap of the hard cores of diameter $D$ is forbidden ($\beta U^+ = \infty$). Note that a connected configuration has $f^+=1$ and disconnected one has $f^+=0$.

Here we also use the Parsons-Lee correction,\cite{parsons,lee} which effectively includes the higher order virial coefficients to make the second-virial theory more accurate for particles with smaller aspect ratios.\cite{schilling2015,meyer2015} This correction consists of using the closure $\hat{C}^+(q,\Omega_{12}) = \Gamma(\phi) \hat{f}^+(q,\Omega_{12})$, where $\Gamma(\phi) = (1-3\phi/4)/(1-\phi)^2$ with packing fraction $\phi = \rho \, v_0$ and $v_0$ the single particle volume. This closure has been shown to give good agreement with simulations for the percolation threshold of moderate aspect ratio hard spherocylinders ($L/D \gtrsim 10$).\cite{schilling2015} Now combining this closure with Eq.~\eqref{eq:percDensity}, we obtain for the percolation packing fraction\cite{schilling2015}
	\begin{equation}\label{eq:percEta}
		\phi_P = \dfrac{2(1+2A-\sqrt{1+A})}{3+4A}
	\end{equation}
with
	\begin{equation}\label{eq:percEta2}
		A = \dfrac{v_0}{  \langle \hat{f}^+(0,\Omega) \rangle_\Omega},
	\end{equation}
where $\hat{f}^+(0,\Omega) = \lim_{q \to 0} \hat{f}^+(q,\Omega)$.

Eq.~\eqref{eq:percEta2} together with the connectedness pair potential in Eq.~\eqref{eq:potential} can be calculated for a fixed particle shape and connectedness criterion $\Delta$. Our approach,\cite{belli2014,dussi2015} relies on Monte Carlo integration of the overlap volume using a large number of two-particle configurations. For all results presented here, we use ten independent runs of $10^9$ Monte Carlo steps, which we found to provide high accuracy even for the largest aspect ratios ($L=100D$) studied, with a typical relative standard error associated with the average overlap volume much smaller than $1\%$. 

\begin{figure}[tbph]
	\centering
			\includegraphics[width=\linewidth]{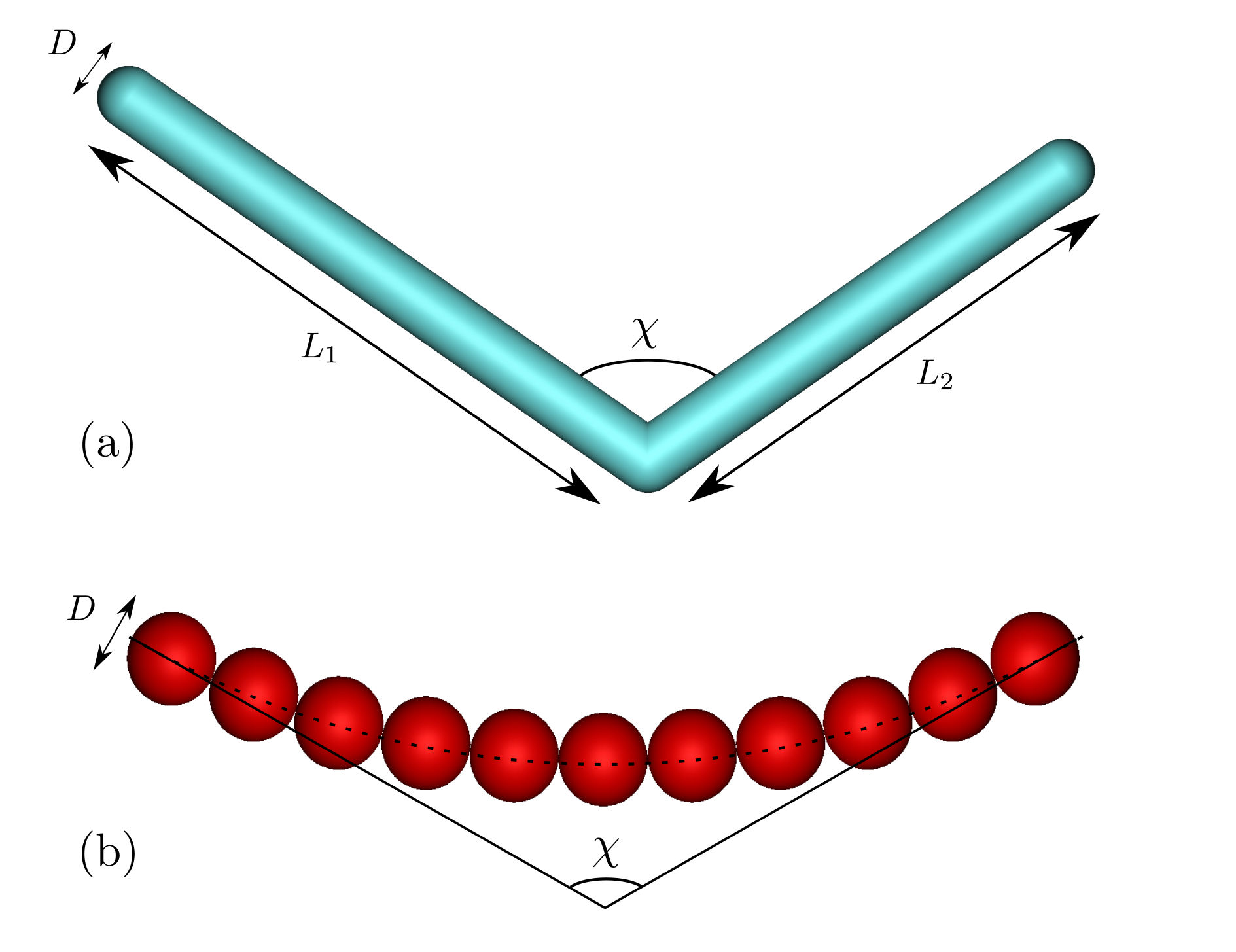}
		\caption{(a) Our model of a kinked rod consisting of two spherocylinders joined at one end with an interarm angle of $\chi$, arm lengths $L_1$ and $L_2$, and diameter $D$. (b) Our model of a bent particle, which consists of $N_s$ fused spheres with diameter $D$ positioned along a circular arc defined by the angle of the end tangents $\chi$.}\label{fig:particleModel}
	\end{figure}

The first particle model we consider is a kinked spherocylinder, shown in Fig.~\ref{fig:particleModel}(a), made up of two spherocylinders of lengths $L_1$ and $L_2$ and identical diameter $D$ which are joined at an angle $\chi$. The second particle, shown in Fig.~\ref{fig:particleModel}(b), models a bent rod, which consists of a set of rigidly fused, tangent beads along a circular arc, with end tangents given by angle $\chi$ and with contour length $L_c = N_s D$, where $D$ is the diameter of the spheres and $N_s$ the number of spheres. 

\begin{figure}[tbph]
	\centering
			\includegraphics[width=\linewidth]{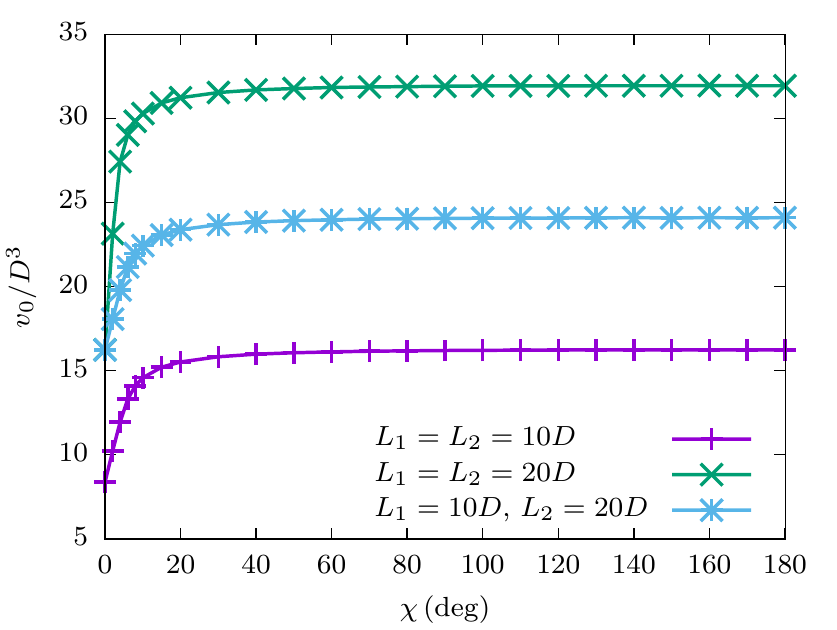}
		\caption{Single particle volume $v_0$ (normalized by diameter $D^3$) of a kinked rod as a function of opening angle $\chi$, with various arm lengths $L_1$ and $L_2$.}\label{fig:boomerangVol}
	\end{figure}

For the kinked rods, the single particle volume decreases slightly as $\chi$ becomes small but nonzero, as the two arms start to intersect. Therefore we have used Monte Carlo integration to determine the single particle volume, which is shown for various arm lengths $L_1$ and $L_2$ as a function of $\chi$ in Fig.~\ref{fig:boomerangVol}.

In Sec.~\ref{sect:results}, we apply connectedness percolation theory to our models of kinked and bent rods, for various particle aspect ratios and deformations $\chi$.

\section{Results}\label{sect:results}

We now show our results for the kinked rod model. Here we are interested in how the percolation packing fraction $\phi_P$ depends on the kink location and kink angle, and if the long-rod scaling is affected. As is the case for straight rods, we expect the percolation packing fraction to also depend on aspect ratio and connectedness criterion $\Delta$. For comparison, we use the analytical form of the percolation packing fraction for straight rods, that is, spherocylinders with length $L$ and diameter $D$, which in the Parsons-Lee second-virial approximation is given by Eq.~\eqref{eq:percEta} with\cite{onsager1949,berhan2007a}
\begin{eqnarray}\label{eq:percRod}
	A =& \, v_0^\text{rod}& \, \left[  \frac{\pi}{2} L^2(D+\Delta) + 2\pi L(D+\Delta)^2 + \frac{4}{3}\pi (D+\Delta)^3 \right. \nonumber\\
	&\quad& - \left. \left( \frac{\pi}{2}L^2 D + 2\pi LD^2 + \frac{4}{3}\pi D^3 \right)  \right]^{-1},\label{eq:percRod}
\end{eqnarray}
with the single particle volume of a spherocylinder $v_0^\text{rod} = \pi LD^2/4+\pi D^3/6$.

\begin{figure}[tbph]
	        \centering
	        \includegraphics[width=\textwidth]{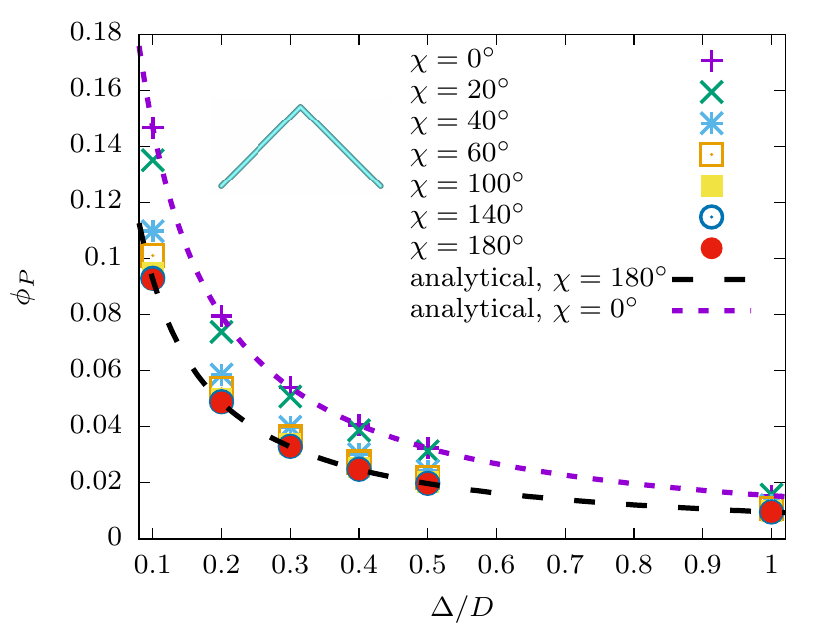}
	        \caption{Percolation packing fraction $\phi_P$ of kinked rods as a function of connectedness criterion $\Delta/D$ for fixed arm lengths $L_1=L_2=20D$ and for various opening angles $\chi$. For comparison, the analytical results for straight spherocylinders are also plotted (dashed curves). Inset illustration shows the particle with $\chi=90^\circ$. }\label{fig:boomerang}
	    \end{figure}

In Fig.~\ref{fig:boomerang}, we show the percolation packing fraction $\phi_P$ as a function of the connectedness criterion (normalized by the rod diameter) $\Delta/D$ for arm lengths $L_1=L_2=20D$ and for various opening angles $\chi$. Here we add the analytical results from Eq.~\eqref{eq:percRod} (dashed curves) for comparison in the two limiting cases of $\chi=180^\circ$, where the kinked rod reduces to a straight rod of length $L=40D$ and $\chi=0^\circ$, where it reduces to a rod of length $L=20D$. First we note that our numerical results are in good agreement with the analytical results from Eq.~\eqref{eq:percRod} in these limiting cases. As in the case of the straight rods, we see that the percolation threshold for kinked rods decreases monotonically with the connectedness criterion $\Delta$.

\begin{figure}[tbph]
	        \centering
	        \includegraphics[width=\textwidth]{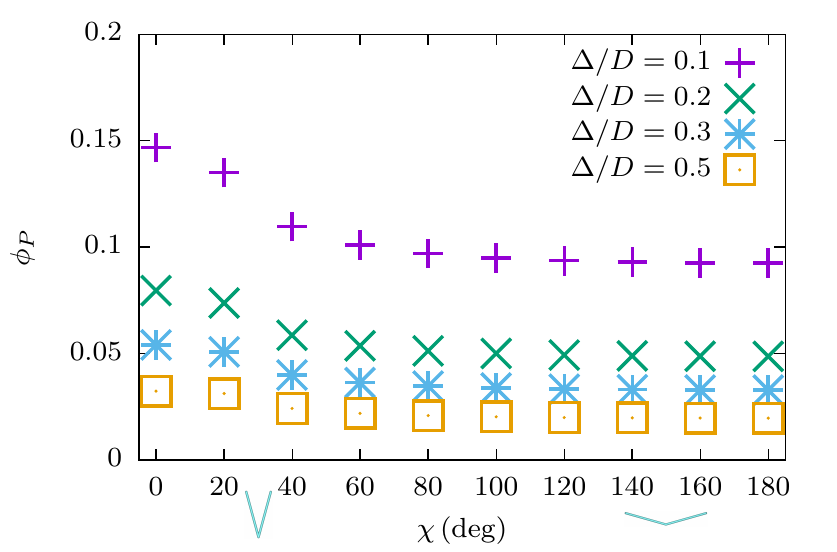}
	        \caption{Percolation packing fraction $\phi_P$ of kinked rods as a function of opening angles $\chi$ for fixed arm lengths $L_1=L_2=20D$ and for various connectedness criteria $\Delta/D$. Illustrations along the horizontal axes show the particle shape for a given angle.}\label{fig:boomerangDelta}
	    \end{figure}

In order to more clearly see the angular dependence, in Fig.~\ref{fig:boomerangDelta} we plot the percolation packing fraction as a function of angle $\chi$ for fixed values of the connectedness criterion $\Delta/D$, again for $L_1=L_2=20D$. Interestingly, we see that for small or even moderate deviations from straight rod shape ($\chi \geq 100^\circ$) there is almost no change (less than a $5\%$ increase) in the percolation threshold. Only at large deformations $\chi \sim 40^\circ$ we see a visible increase in the percolation threshold, which becomes on the order of a $50\%$ increase for $\chi=20^\circ$. We can explain this increase in the percolation threshold as due to a decrease in available connected volume, since the effective aspect ratio of the particles is significantly decreased. Going from $\chi=20^\circ$ to $\chi=0^\circ$, there are two competing effects: first that the aspect ratio is further decreased and so $\phi_P$ increases, and second that the single-particle volume decreases as the particle arms overlap (see Fig.~\ref{fig:boomerangVol}), which decreases $\phi_P$.

\begin{figure*}[tbph]
	        \centering
	        \includegraphics[width=\textwidth]{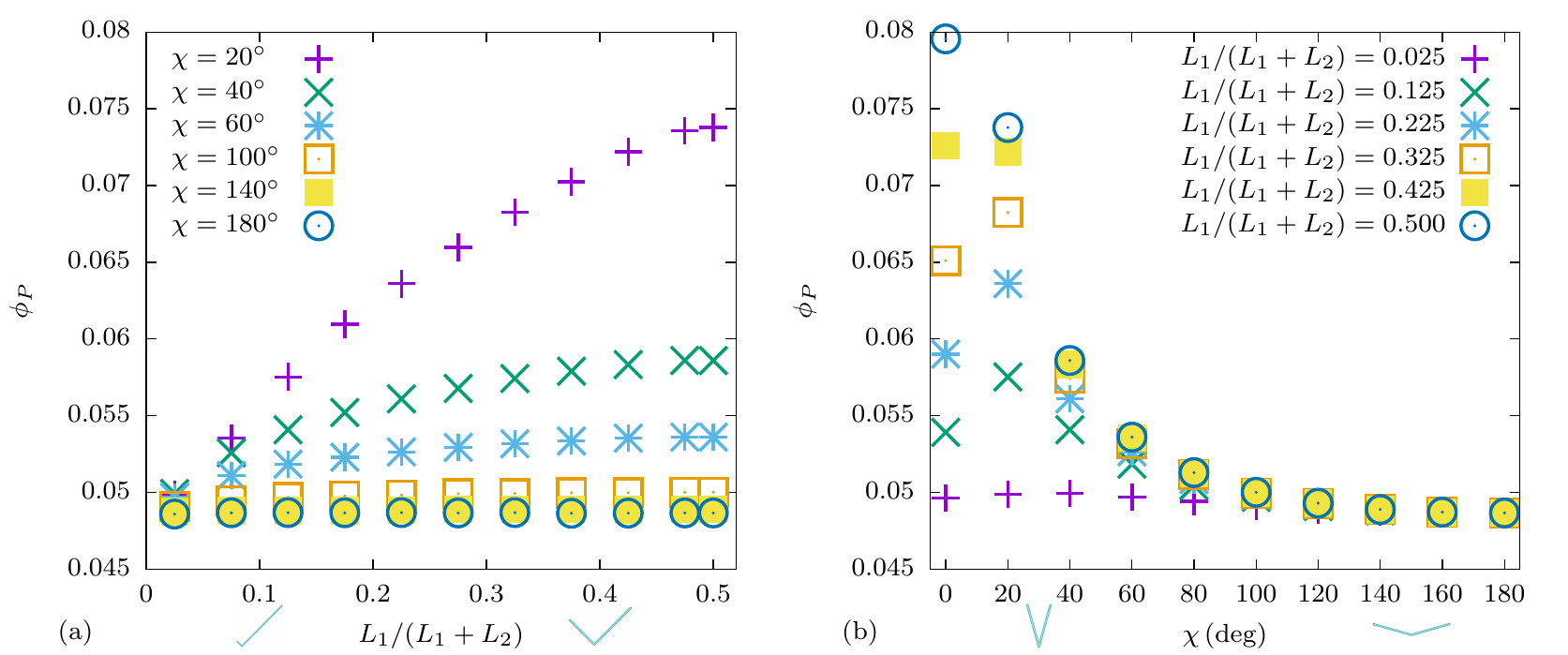}
	        \caption{Percolation packing fraction $\phi_P$ of kinked rods with fixed total length $L_1+L_2=40D$ and connectedness criterion $\Delta/D=0.2$, (a) as a function of kink position ($L_1/(L_1+L_2)$) for various kink angles $\chi$ and (b) as a function of kink angle for various kink positions. Illustrations along the horizontal axis show (a) the corresponding particle shape for $\chi=90^\circ$ and (b) the variation of the angle for $L_1/(L_1+L_2)=0.5$.  }\label{fig:boomerangFixedL}
	    \end{figure*}

Next we study the dependence on kink location, for a fixed total rod length of $L_1+L_2=40D$ and a fixed connectedness criterion $\Delta/D=0.2$. In Fig.~\ref{fig:boomerangFixedL}(a), we show the percolation packing fraction $\phi_P$ vs.\ the kink location $L_1/(L_1+L_2)$ for various angles $\chi$. For kink location $L_1/(L_1+L_2) \approx 0$, as expected, $\phi_P$ cannot depend on angle $\chi$. In fact, we see that the greatest deviation from straight rod behavior is for a central kink ($L_1/(L_1+L_2) =0.5$). In  Fig.~\ref{fig:boomerangFixedL}(b), we plot the percolation threshold as a function of the kink angle $\chi$ for different values of the kink location $L_1/(L_1+L_2)$. This illustrates again that for small deformations $\chi \lesssim 180^\circ$, there is very little effect on the percolation threshold $\phi_P$. The percolation threshold increases as the kink angle decreases towards $\chi \approx 20^\circ$, which in the case of a central kink is about a $50\%$ increase compared with $\chi=180^\circ$. The maximum in the percolation threshold for some kink locations in Fig.~\ref{fig:boomerangFixedL}(b) is caused by the decrease in the single-particle volume between $\chi=20^\circ$ and $\chi=0^\circ$ (see Fig.~\ref{fig:boomerangVol}), which in turn decreases the percolation threshold.

\begin{figure*}[tbph]
	        \centering
	        \includegraphics[width=\textwidth]{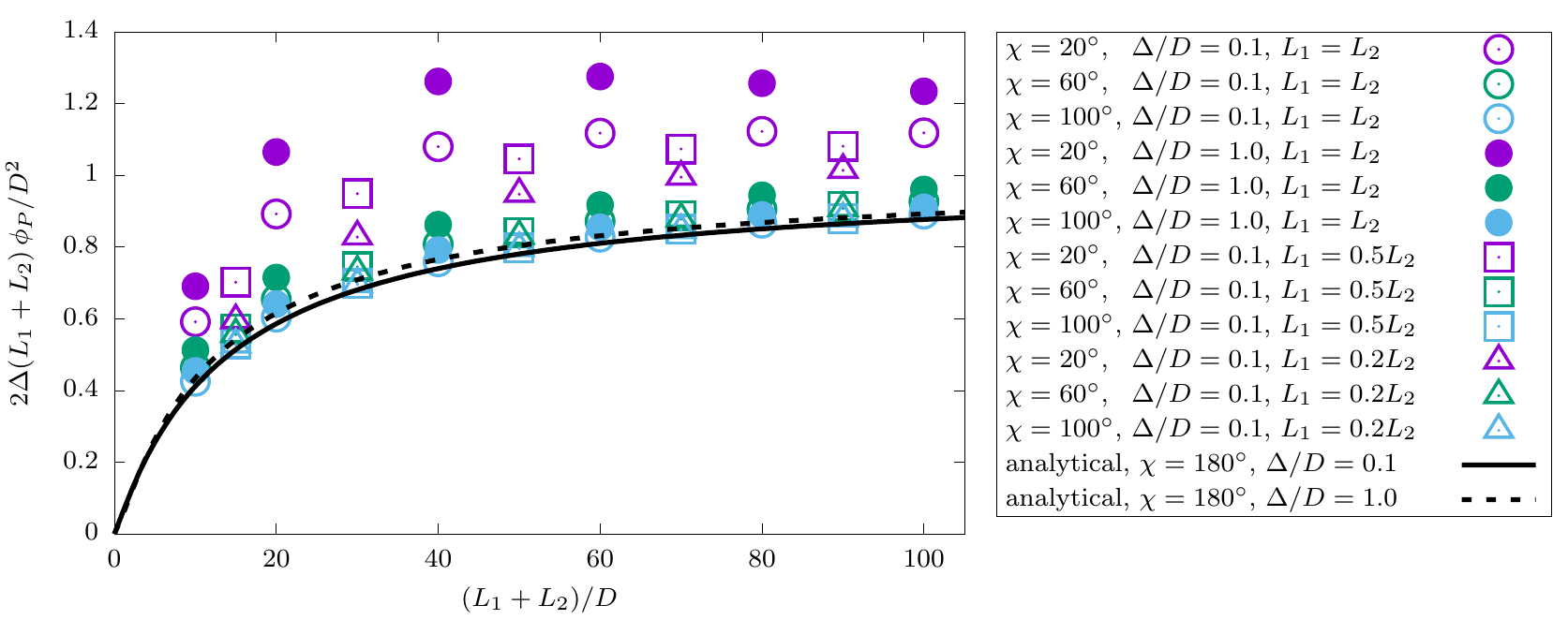}
	        \caption{Percolation packing fraction $\phi_P$ (scaled by $2\Delta(L_1+L_2)/D^2$) as a function of total length $(L_1+L_2)/D$ for kinked rods with arms $L_1=L_2$ (circles), $L_1=0.5L_2$ (squares), and $L_1=0.2L_2$ (triangles), with connectedness criterion $\Delta/D =0.1$ (empty symbols) and $\Delta/D = 1.0$ (filled symbols), and for opening angles $\chi=20^\circ$ (purple), $\chi=60^\circ$ (green), and $\chi=100^\circ$ (blue). The curves show analytical results for straight rods of length $L_1+L_2$, with $\Delta/D=0.1$ (solid) and $\Delta/D=1.0$ (dashed). }\label{fig:boomerangL}
	    \end{figure*}

Finally, we want to examine the shape dependence of the large aspect ratio scaling behavior. As we can see from Eq.~\eqref{eq:percRod}, for straight rods in the limit $L \gg D,\Delta$, we have that $\phi_P \to D^2/(2L\Delta)$. Therefore we multiply $\phi_P$ by  $2L\Delta/D^2$, such that for straight rods it approaches unity in the large aspect ratio limit. In Fig.~\ref{fig:boomerangL}, we show the scaled percolation thresholds as a function of the aspect ratio for many parameters in one plot. Here the kink angle dependence is shown by the colors: purple ($\chi=20^\circ$), green ($\chi=60^\circ$), and blue ($\chi=100^\circ$). We also vary the connectedness criterion and the kink location, with $\Delta/D=0.1$ given by the empty symbols and $\Delta/D=1.0$ given by the filled symbols, and with the circles representing a central kink ($L_1=L_2$) and the squares and triangles showing two asymmetric cases ($L_1=0.5L_2$ and $L_1=0.2L_2$ respectively). For comparison we plot the analytical results for straight rods of length $L_1+L_2$ with $\Delta/D=0.1$ (solid) and $\Delta/D=1.0$ (dashed). Strikingly, we see that even for relatively large deformations of up to $\chi = 60^\circ$, there is only a very small deviation from straight-rod asymptotic behavior, not exceeding $10\%$ for $L_1+L_2 \geq 80D$, which is true for any kink location or connectedness criteria. Only in the most extreme deformation considered here, $\chi=20^\circ$, do we see larger deviations, on the order of $35\%$ as $(L_1+L_2)/D$ becomes large. For these small angles $\chi$, some $\chi$-dependent corrections to the straight-rod scaling must be important. Of course, this is to be expected since the relevant aspect ratio is no longer $(L_1+L_2)/D$ when $\chi$ becomes small. We also note that this extreme case of $\chi=20^\circ$ is most likely not the most relevant case for real experimental systems.

Now we consider the second model described in Sec.~\ref{sect:method}, namely of a bent rod, modeled by a bead chain along a circular arc (see Fig.~\ref{fig:particleModel}(b)). As we have already discussed in detail the effect of varying the aspect ratio for the kinked rod model, here we restrict ourselves to varying $\chi$ for a bent rod consisting of $N_s=11$ tangent spheres of diameter $D$, with contour length $L_c=11D$. The bend angle $\chi$ (as illustrated in Fig.~\ref{fig:particleModel}(b)) can vary from $\chi=180^\circ$ (straight rod) to $\chi=0^\circ$ (a half circle). 

\begin{figure}[tbph]
	        \centering
	        \includegraphics[width=\textwidth]{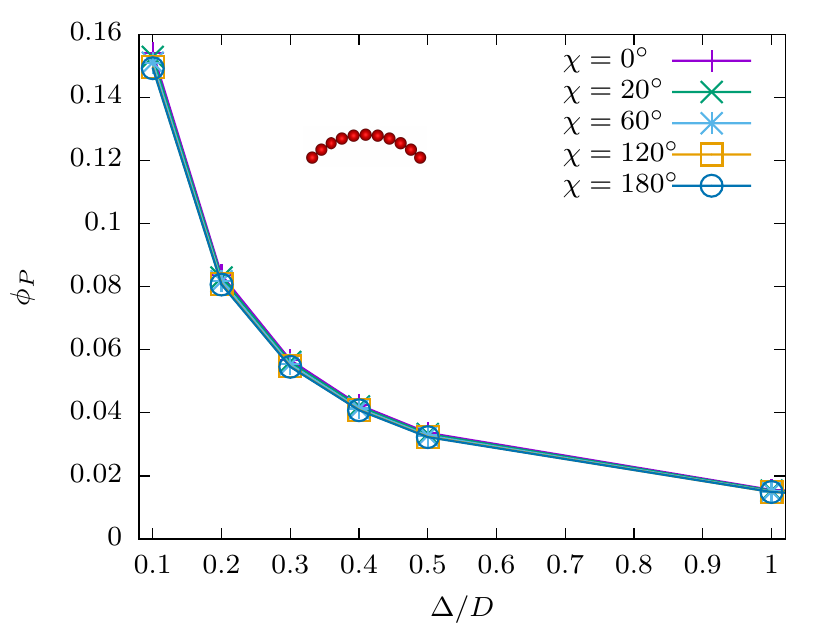}
	        \caption{Percolation packing fraction $\phi_P$ of bent rods with fixed number of spheres $N_s=11$ as a function of connectedness criterion $\Delta/D$ for various bend angles $\chi$. Inset illustration shows the particle shape for $\chi=90^\circ$.}\label{fig:crescentPlot}
	    \end{figure}

\begin{figure}[tbph]
	        \centering
	        \includegraphics[width=\textwidth]{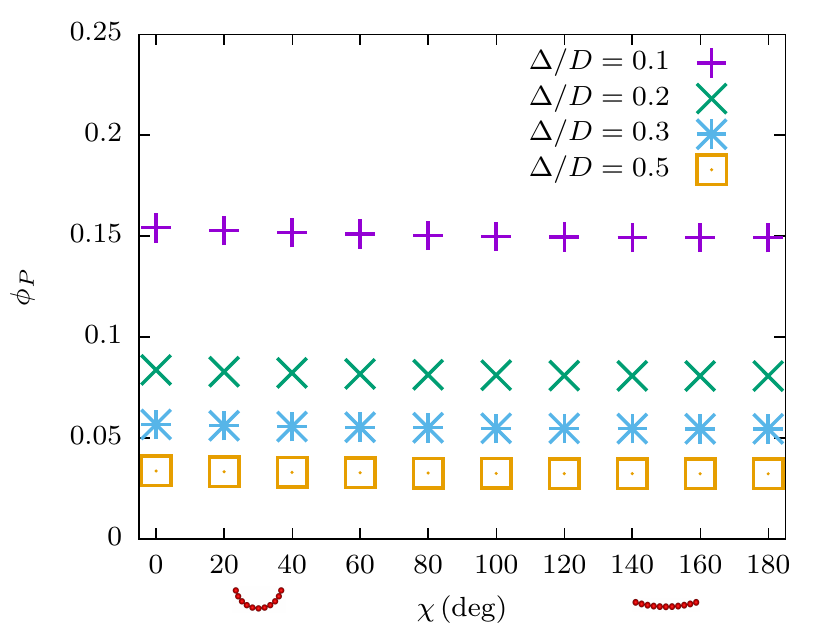}
	        \caption{Percolation packing fraction $\phi_P$ of bent rods with fixed number of spheres $N_s=11$ as a function of bend angle $\chi$ for various connectedness criteria $\Delta/D$. Illustrations along the x-axis show the particle shape for a given angle.}\label{fig:crescentDelta}
	    \end{figure}

As before, we first consider the percolation packing fraction $\phi_P$ as a function of connectedness criterion $\Delta$ for various angles $\chi$ (Fig.~\ref{fig:crescentPlot}). As for a straight rod, $\phi_P$ decreases monotonically with increasing $\Delta$, for all $\chi$. We plot in Fig.~\ref{fig:crescentDelta} the percolation threshold as a function of the bend angle $\chi$, for various $\Delta/D$ and see that deforming a straight rod ($\chi=180^\circ$) into a half circle ($\chi=0^\circ$) has no visible effect at all on the percolation threshold. This suggests that bending fluctuations also have a very small effect on the percolation threshold.\cite{kyrylyuk2008} This result is consistent with the behavior of the kinked rods as they vary from $\chi=180^\circ$ to $\chi=90^\circ$. We emphasize that the two particle shapes have different definitions of $\chi$ as shown in Fig.~\ref{fig:particleModel}, with the bent particles being less deformed at $\chi=0^\circ$, where they are more comparable in shape to a kinked rods with $\chi=90^\circ$. In the following section we give a summary of our findings and an outlook on future research directions.

\section{Discussion and Conclusions}\label{sect:conclusions}

In this paper, we have used connectedness percolation theory with the Parsons-Lee second-virial closure to study kinked and bent rodlike nanofillers. We calculated the percolation threshold, which is inversely proportional to an average overlap volume, using Monte Carlo integration. We have shown that the percolation threshold is only very weakly affected by small or even moderate rod shape deformations. For larger deformations, we saw a small increase in the percolation threshold. In addition, the universal scaling with particle aspect ratio and connectedness criterion was only affected for very deformed particles, which can be seen as due to an effective reduction in the aspect ratio.

Our approach of combining connectedness percolation theory with Monte Carlo integration is able to deal with any complicated particle shape provided that one has a two-particle overlap algorithm. It is exact in the large aspect ratio limit and since it uses the Parsons-Lee correction, we also expect it to be reasonably accurate for moderate aspect ratios, though more work is needed to understand this correction's applicability to non-rodlike particle shapes. We note that the only effect of the Parsons-Lee correction is to shift the percolation threshold to lower packing fractions. The qualitative behavior we find is completely unchanged by adding this correction.

Although previous works have not considered the explicit dependence on a kink or bend deformation, several studies have dealt with the effect of waviness on the percolation threshold of long rods. Based on the theory of fluids of flexible rods, it has been predicted that a finite bending flexibility weakly increases the percolation threshold since the bending effectively decreases the length.\cite{kyrylyuk2008} There have been several results from simulations that also find a weak increase in the percolation threshold due to flexibility or waviness.\cite{berhan2007b,li2008,dalmas2006} Notably, in Ref.~[\citen{berhan2007b}], randomly oriented wavy fibers with different curvatures were studied through both simulations and excluded volume calculations. Here it was found that in the large aspect ratio limit, the percolation threshold of wavy rods was comparable to but slightly larger than that of straight rods.\cite{berhan2007b}

The fact that moderate kinks in long rods do not affect their percolation threshold can be understood qualitatively by an argument similar to the one given in Ref.~[\citen{khokhlov1981}] and also the basis for a common approximation used in, e.g., Refs.~[\citen{teixeira1998,blaak1998}]. Recall that in the long-rod limit $L \gg D$ and $L\gg \Delta$, the inverse percolation density, which is also the connectedness version of the average excluded volume, is $\rho_P^{-1} = \langle \hat{f}^+(0,\Omega) \rangle_\Omega = \pi L^2 \Delta/2$. Suppose that we consider each rod as consisting of two segments of length $L/2$, core diameter $D$, and shell diameter $D+\Delta$, then the average excluded volume can be written as the sum of the average segmentwise excluded volumes $\rho_P^{-1} = 4 \pi (L/2)^2 \Delta/2 $. This yields the same result as for the original rods and is still exact in the long rod limit, since we ignore end effects. Now, consider that the two segments of the rods are joined at some angle $\chi$. As before, we can write the excluded volume as a sum of the segmentwise excluded volumes, which implies that the percolation threshold is the same as for a straight rod. However, this is no longer exact, and in fact it becomes a worse approximation as the rods become more deformed. This is because it becomes more probable for two rods to have a simultaneous overlap of both pairs of segments and so the segmentwise excluded volume overestimates the true overlap volume.\cite{bisi2008} Therefore this qualitative argument only applies to moderately deformed rods.

In the future it would be interesting to examine the structures of the clusters of kinked and bent rods, as well as their percolation thresholds in the prolate, oblate, and biaxial nematic phases. Also, mixtures of deformed particles or polydisperse systems with defects would be an interesting future investigation.


\section*{Acknowledgments}

This work is part of the D-ITP consortium, a program of the Netherlands Organization for Scientific Research (NWO) that is funded by the Dutch Ministry of Education, Culture and Science (OCW). We also acknowledge financial support from an NWO-VICI grant.

\bibliography{percolation}

\end{document}